\documentclass[aps,prl,twocolumn,preprintnumbers,amsmath,amssymb,showpacs]{revtex4-1}
\usepackage{amsmath}
\usepackage{graphicx}
\usepackage{amssymb}
\usepackage{color}
\usepackage{dcolumn}
\usepackage{epsfig}
\usepackage{bm}
\begin{document}

\title{Optimized contraction scheme for tensor-network states}

\author{Z. Y. Xie$^1$}
\author{H. J. Liao$^{2}$}
\author{R. Z. Huang$^{2, 3}$}
\author{H. D. Xie$^{2, 3}$}
\author{J. Chen$^{2, 3}$}
\author{Z. Y. Liu$^{3, 4}$}
\author{T. Xiang$^{2, 5}$}
\email{txiang@iphy.ac.cn}

\affiliation{$^1$Department of Physics, Renmin University of China, Beijing 100872, China}

\affiliation{$^2$Institute of Physics, Chinese Academy of Sciences, Beijing 100190, China}

\affiliation{$^3$School of Physical Sciences, University of Chinese Academy of Sciences, Beijing 100049, China}

\affiliation{$^4$Institute of Theoretical Physics, Chinese Academy of Sciences, Beijing 100190, China}

\affiliation{$^5$Collaborative Innovation Center of Quantum Matter, Beijing 100190, China}

\date{\today}

\begin{abstract}
  In the tensor-network framework, the expectation values of two-dimensional quantum states are evaluated by contracting a double-layer tensor network constructed from  initial and final tensor-network states.
  The computational cost of carrying out this contraction is generally very high, which limits the largest bond dimension of tensor-network states that can be accurately studied to a relatively small value.
  We propose an optimized contraction scheme to solve this problem by mapping the  double-layer tensor network onto an intersected single-layer tensor network.
  This reduces greatly the bond dimensions of local tensors to be contracted, and improves dramatically the efficiency and accuracy of the evaluation of expectation values of tensor-network states.
  It almost doubles the largest bond dimension of tensor-network states whose physical properties can be efficiently and reliably calculated, and it extends significantly the application scope of tensor-network methods.
\end{abstract}

\maketitle

\section{Introduction}


Strongly correlated quantum spin or fermion systems pose some most intriguing problems, which are difficult to solve due to the exponential growth of the dimension of Hilbert space with the system size and the lack of small parameters that can be used to carry out perturbative calculations in quantum field theory.
Numerical simulations have emerged as an indispensable tool and achieved immense progress in recent years.
The quantum Monte Carlo \cite{QMCOld, QMCNew} and density matrix renormalization group (DMRG) \cite{DMRG1992} are the two most commonly used numerical methods in the study of quantum lattice models.
The quantum Monte Carlo is a powerful method for studying interacting bosons, quantum spin models without frustrations, and some special interacting fermion models. However, it suffers from the so-called minus-sign problem~\cite{LW2015, MajPos2016} in dealing with interacting fermions or frustrated quantum spin models in which the error increases exponentially with the system size and with decreasing temperature.
The DMRG is the most accurate method for studying one-dimensional systems, but in two or higher dimensions, the lattice size that can be reliably handled by the DMRG \cite{XiangPRB2001,DMRG2D} is relatively small, limited by the entanglement area law \cite{AreaLaw} which implies that the number of states retained must increase exponentially with the boundary size.

In recent years, a new class of numerical methods which combine the renormalization group techniques~\cite{CTM1, TEBD2, iTEBD1, JiangPRL2008, TRG, SRG, HHZhao2010, HOTRG} with the  tensor-network representation of quantum many-body states or partition functions of statistical models~\cite{TPS1, TPS2, TPS3, PEPS2004, MERA, PESS2014} has  been developed.
These methods have played an important role in the study of strongly correlated problems, especially in two or higher dimensions~\cite{HOTRG, CTMfpeps, LiaoPRL2017}.
A tensor-network state is a variational ansatz for the ground-state wave function with embedded entanglement structures~\cite{TPS1, TPS2, TPS3, PEPS2004, MERA, PESS2014}.
It is a generalization of the one-dimensional matrix product state (MPS), which is the wave function generated by the DMRG calculation~\cite{DMRGandMPS, MPSandTree}.
Among the various kinds of constructions, the projected entangled pair state (PEPS)~\cite{PEPS2004} and, more generally, the projected entangled simplex state (PESS)~\cite{PESS2014} are two examples of tensor-network states that are commonly used.
Both represent faithfully the ground states of quantum lattice models that satisfy the area law of entanglement entropy.

The accuracy of a tensor-network state is controlled by the bond dimension $D$ of local tensors, which determines the number of variational parameters.
In one dimension, $D$ is just the number of basis states retained in the DMRG calculation.
In order to obtain accurate results for the ground-state energy and other physical quantities, one should keep $D$ as large as possible and check the convergence
behavior of the results with increasing $D$.
However, the computational cost increases very rapidly with an increase in $D$ in two dimensions.
This has limited the largest $D$ that can be handled using the tensor-network algorithms currently available for the PEPS or PESS to generally less than 14 \cite{PESS2014}.

\begin{figure*}[t]
\begin{center}
\includegraphics[width=15cm]{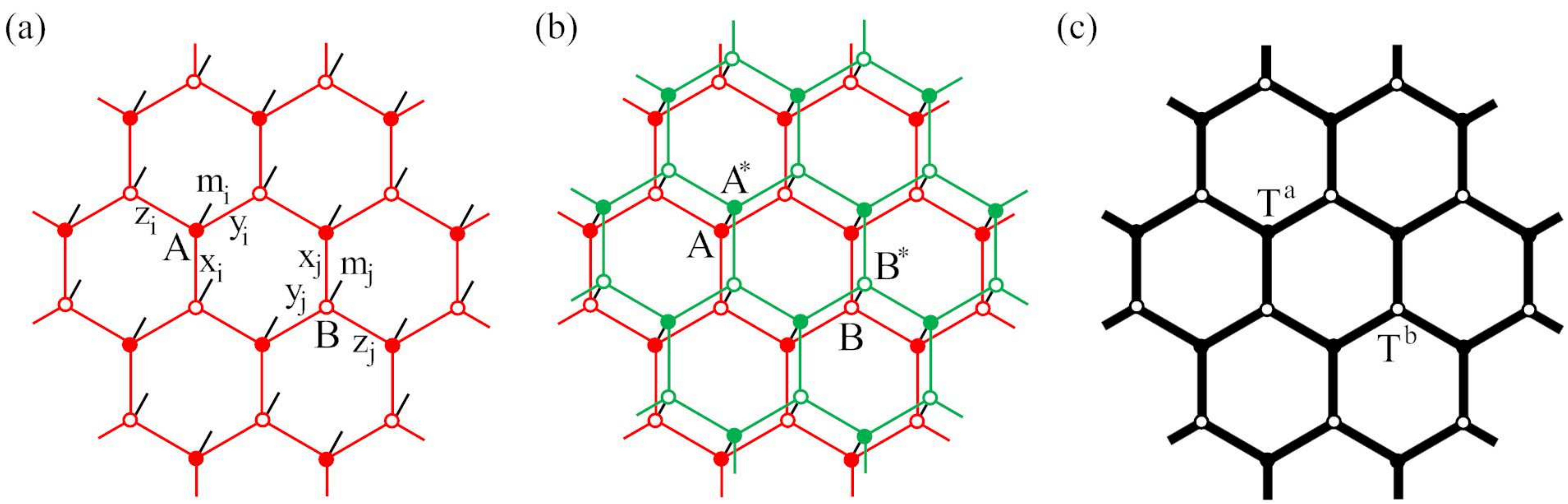}
\end{center}
  \caption{ (a) PEPS wave function, $|\Psi\rangle$, defined on the honeycomb lattice. $A_{x_iy_iz_i}$ and $B_{x_jy_jz_j}$ are the local tensors defined on sublattices $A$ and $B$, respectively. The dangling bond, $m_i$, represents a physical basis state at site $i$. $x_i$, $y_i$, and $z_i$ are the bond variables of dimension $D$ linking site $i$ along the $x$, $y$, and $z$ directions, respectively.
  (b) Double-layer tensor-network representation of $\langle \Psi| \Psi \rangle$. The red and green layers, linked by black physical bonds, represent $|\Psi\rangle$ and $\langle \Psi |$, respectively.
  (c) The reduced single-layer tensor network, $\langle\Psi|\Psi\rangle$, whose bond dimensions are all equal to $D^2$, obtained from (b) by tracing out all physical bonds. }
\label{Lattice}
\end{figure*}

To understand this, let us take the translational-invariant PEPS defined on the honeycomb lattice (or a PESS defined on a Kagome lattice) shown in Fig.~\ref{Lattice}(a), i.e.,
\begin{equation}
  |\Psi\rangle = \mathrm{Tr}\prod_{i\in A ,j\in B} A_{x_iy_iz_i}[m_i]B_{x_jy_jz_j}[m_j] |m_im_j \rangle , \label{eq:PEPSwf}
\end{equation}
as an example to explain why the cost is so high. In Eq.~(\ref{eq:PEPSwf}), $A_{x_iy_iz_i}[m_i]$ and $B_{x_jy_jz_j}[m_j]$ are the local tensors defined in sublattices A and B of the honeycomb lattice, respectively.
$m_i$ is the physical basis state at sites $i$ and $(x_i,y_i,z_i)$ are the three virtual bond variables linking that site.
The trace is to sum over all physical states  and over all virtual bond states.
The elements of local tensors $A$ and $B$ are variational parameters.
They can be approximately determined, for example, through an imaginary-time evolution by taking an entanglement mean-field approximation (also called the simple update method in the literature)~\cite{JiangPRL2008}.
With this approach, a PEPS whose bond dimension is as large as 100 or more can be readily calculated \cite{Husimi2016}.
However, to calculate physical propreties, one needs to evaluate the expectation values of physical observables $\hat{O}$ using the formula,
\begin{equation}
\langle O\rangle = \frac{\langle\Psi|\hat{O}|\Psi\rangle}{\langle\Psi|\Psi\rangle}  . \label{eq:Exp}
\end{equation}
Unfortunately, this part of the calculation is not so efficient. As discussed below, its cost scales as $D^{12}$ in the large-$D$ limit for PEPS.
This is because, as shown in Fig.~\ref{Lattice}(b), both the numerator and the denominator on the right-hand side of Eq.~(\ref{eq:Exp}) are double-layer tensor networks.
Due to the existence of loops, there is no exact method for contracting these double-layer tensor-network states.
A commonly used approach is, first, to compress this double-layer tensor-network state into a single-layer one [Fig.~\ref{Lattice}(c)] by tracing out all physical degrees of freedoms, i.e., short black bonds in Fig.~\ref{Lattice}(b), and, then, to contract this reduced tensor network (RTN) approximately using the transfer matrix renormalization~\cite{TMRG1, CTM1, TMRG2, TMRG3}, the time-evolving block decimation (TEBD) \cite{TEBD2,iTEBD1}, or other coarse-graining tensor renormalization group algorithms~\cite{TRG, SRG, HHZhao2010, HOTRG}.
This contraction scheme is simple to implement.
However, as discussed in Sec.~\ref{sec:RTN}, its cost is very high because the bond dimension of this RTN is $D^2$, which is significantly larger than the original tensor-network state.
In general, the cost of carrying out this contraction scales as $O(D^8)$ in storage memory and as  $O(D^{12})$ in computational time.
This has limited the maximal accessible $D$ to generally less than or equal to $13$ using the computer resources currently available~\cite{PESS2014}.
In view of the fact that the more accurate determination of the ground-state wave function through higher-cost-demanding techniques, e.g., the full update method~\cite{FU2008, FU2010}, also involves this kind of contraction, it is clear that this high cost of the contraction is the main bottleneck that limits the application of almost all tensor-network algorithms.

\begin{figure}[t]
\begin{center}
\includegraphics[width=7.5cm]{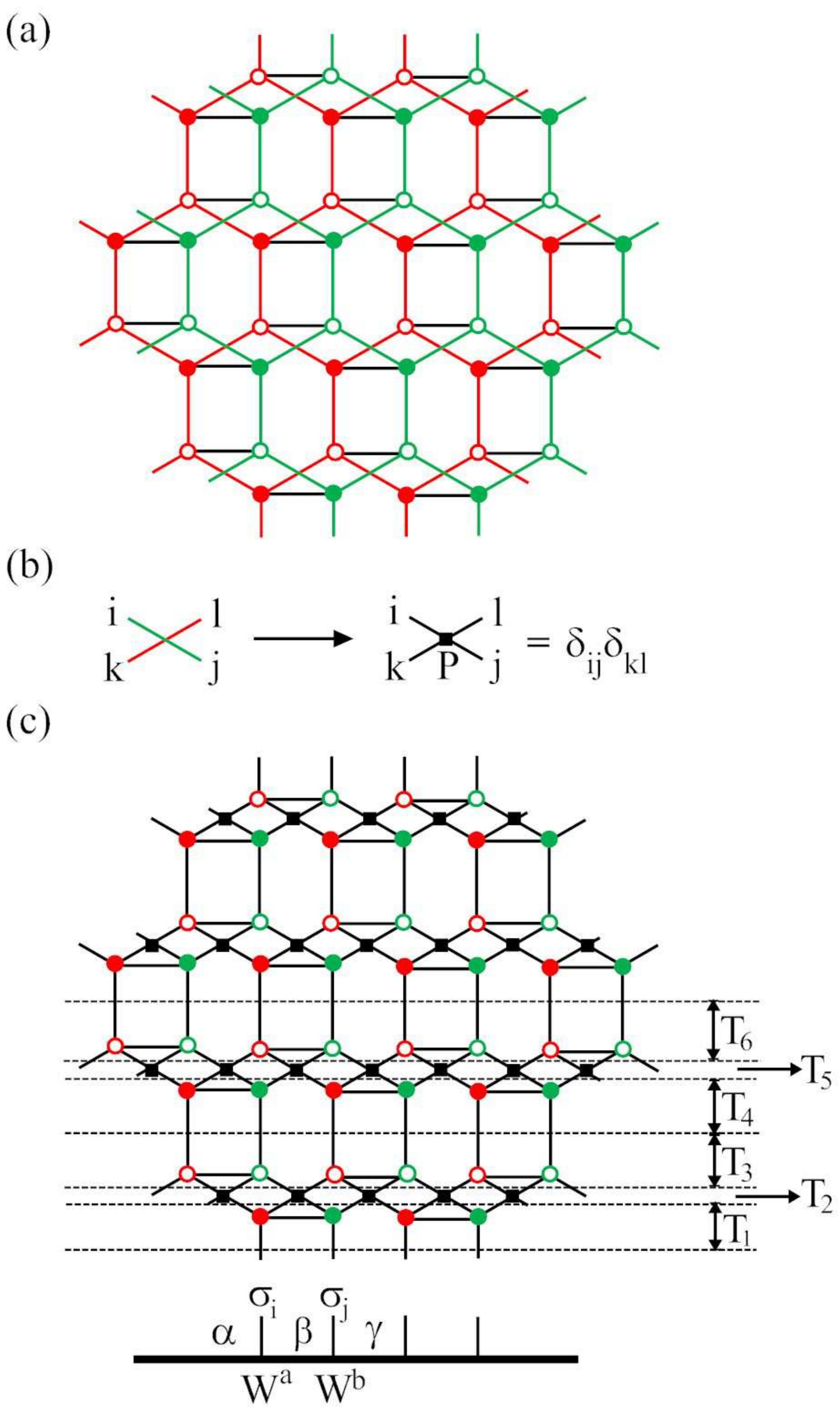}
\end{center}
  \caption{(a) Nested single-layer mapping of the double-layer tensor-network state shown in Fig.~\ref{Lattice}(b) done by shifting the top layer by half a unit cell along the horizontal direction and then compressing it to the bottom layer.
  (b) Virtual swap tensor $P$, which is defined as the direct product of two Kronecker delta functions.
  (c) Resulting NTN (top) and graphical representation of the boundary MPS defined by Eq.~(\ref{eq:BMPS2}) (bottom).
  The transfer matrix $T$ is the product of six sub-transfer matrices $T_i$ ($i= 1,\cdots , 6)$. }
  \label{fig:NTN}
\end{figure}

To solve the above problem, we propose an optimized contraction scheme to transform the double-layer tensor network shown in Fig.~\ref{Lattice}(b) into a single-layer one.
Instead of compressing the two layers into one by contracting the physical bonds, we propose shifting the top layer in Fig.~\ref{Lattice}(b) by half a unit cell along the $x$ axis and then compressing the top layer onto the bottom layer.
This allows us to obtain a single-layer tensor network defined on a lattice of two intersecting honeycomb lattices connected by the physical bonds.
For convenience in the discussion below, we call it a nested tensor network (NTN).
A schematic of this NTN, which is discussed in detail in Sec.~\ref{sec:NTN}, is presented in Fig.~\ref{fig:NTN}(a).
The bond dimensions of the NTN are $D$ on the original virtual bonds and $d$ on the original physical bonds [black bonds in Fig.~\ref{fig:NTN}(a)].
Both are much smaller than the bond dimensions of the corresponding RTN.
As explained in Sec. \ref{sec:NTN}, this reduces the computational time from $O(D^{12})$ to $O(D^9)$ and the memory space from $O(D^8)$ to $O(D^6)$.
The contraction of this NTN can therefore be done much more efficiently.
This can reduce significantly the computational cost and enlarge the maximal accessible $D$ using the existing tensor-network contraction methods.

Another way to solve this problem is to use only $|\Psi\rangle$ or $\langle \Psi |$, instead of their inner-product, to contract a two-dimensional tensor-network state, in combination with the Monte Carlo simulation~\cite{LW2011, WYLiu2016, HHZVMC}.
This can also reduce the cost of the evaluation of expectation values.
However, this method works only on finite lattice systems, and its cost is also quite high because this method requires the sampling of many different physical configurations in order to carry out the Monte Carlo simulation.

The paper is organized as follows.
In Sec.~\ref{sec:RTN}, we introduce briefly the conventional contraction method based on the RTN.
In Sec.~\ref{sec:NTN}, we discuss in detail how to construct and contract an NTN, using a PEPS defined on a honeycomb lattice as an example.
In Sec.~\ref{sec:result}, we present some benchmark results obtained with the NTN method for the antiferromagnetic Kagome Heisenberg model.
We summarize in Sec.~\ref{sec:summary}.

\section{The RTN method} \label{sec:RTN}

Let us first consider how the expectation value of the PEPS, defined in Eq.~(\ref{eq:PEPSwf}), is evaluated in the conventional tensor-network scheme.
In the below, we demonstrate only how the denominator in Eq.~(\ref{eq:Exp}), i.e., $\langle \Psi | \Psi \rangle$, is contracted.
A similar technique can be used to contract the numerator $\langle \Psi | \hat{O} |\Psi \rangle$.

In the conventional contraction scheme, the first step is to compress the double-layer tensor-network state shown in Fig.~\ref{Lattice}(b) into a single-layer one [Fig.~\ref{Lattice}(c)].
This defines a new tensor-network state on the original honeycomb lattice whose local tensors are defined by
\begin{eqnarray}
T^{a}_{xx',yy',zz'} & =& \sum_{m}A_{xyz}[m]A^*_{x'y'z'}[m] , \label{eq:RTN1}\\
T^{b}_{xx',yy',zz'} & =& \sum_{m}B_{xyz}[m]B^*_{x'y'z'}[m] . \label{eq:RTN2}
\end{eqnarray}
The bond dimensions of these tensors equal $D^2$.
As there are no dangling physical bonds, this TNS can be regarded as a tensor-network representation of a classical partition function.
Thus the renormalization group methods that have been developed for contracting the partition functions of classical statistical models, such as the corner transfer matrix renormalization group~\cite{CTM1}, the TEBD~\cite{TEBD2, iTEBD1}, and the coarse-graining tensor renormalization group methods~\cite{TRG, SRG, HHZhao2010, HOTRG}, can be used to contract this tensor network approximately.
Among them, the corner transfer matrix renormalization group and the TEBD are more commonly used.
Their computational costs are of the same order.
For noncritical systems, they are also generally the more efficient, hence less costly, methods for contracting classical partition functions.
Nevertheless, the coarse-graining tensor renormalization group methods~\cite{TRG, SRG, HOTRG} are more convenient to use for analyzing the critical behavior of tensor-network states.

\begin{figure}[t]
\begin{center}
\includegraphics[width=6.5cm]{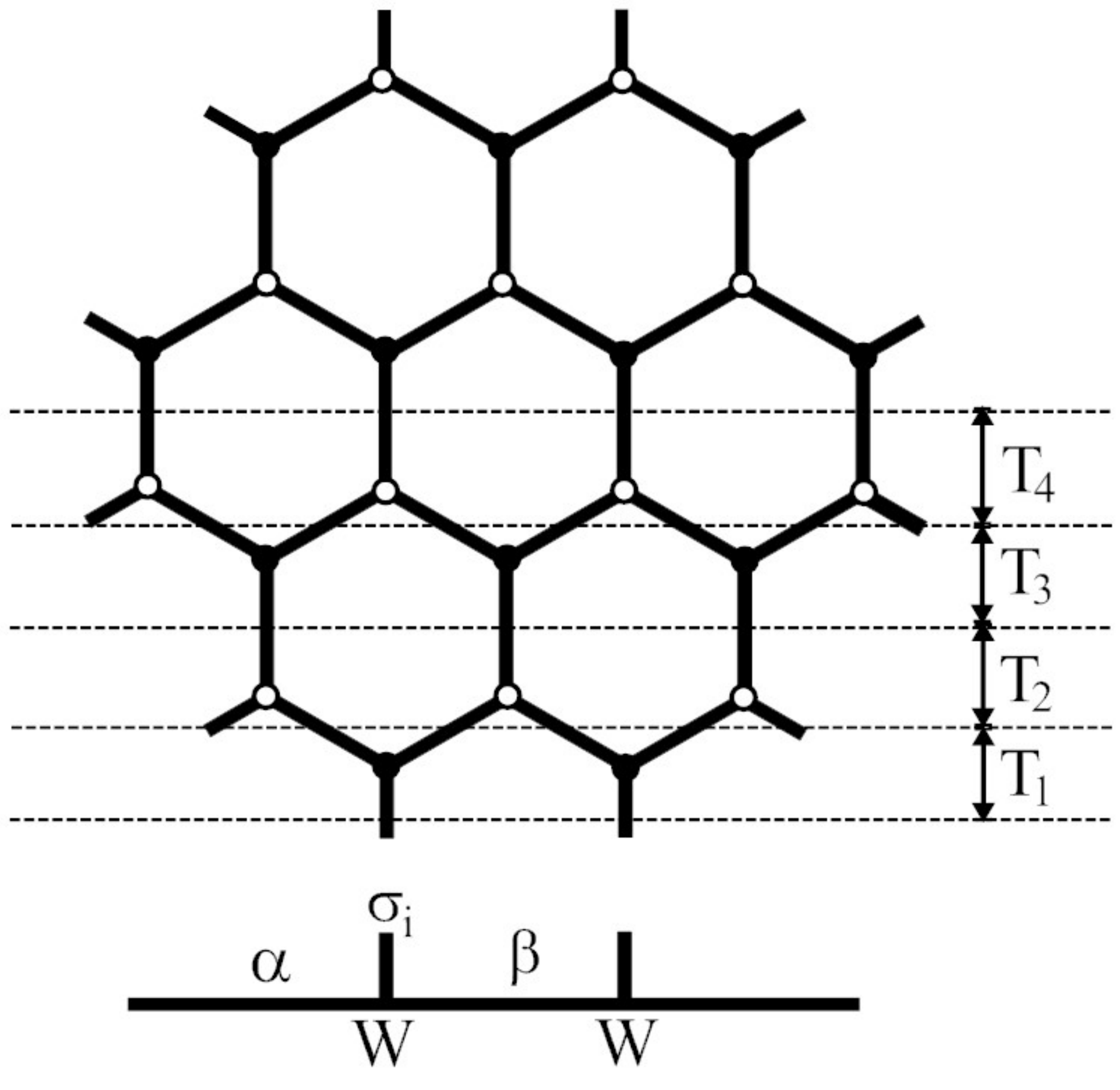}
\end{center}
  \caption{Graphical representation of an RTN whose local tensors are defined by Eqs.~(\ref{eq:RTN1}) and (\ref{eq:RTN2}) (top) and the boundary MPS defined by Eq~(\ref{eq:MPS1}) (bottom). Dashed lines separate the transfer matrices used in the iTEBD calculation. $W_{\alpha\beta}[\sigma_i]$ is the local tensor of the boundary MPS.}
\label{ReducedTN}
\end{figure}

\begin{figure}[tbp]
\begin{center}
\includegraphics[width=6.5cm]{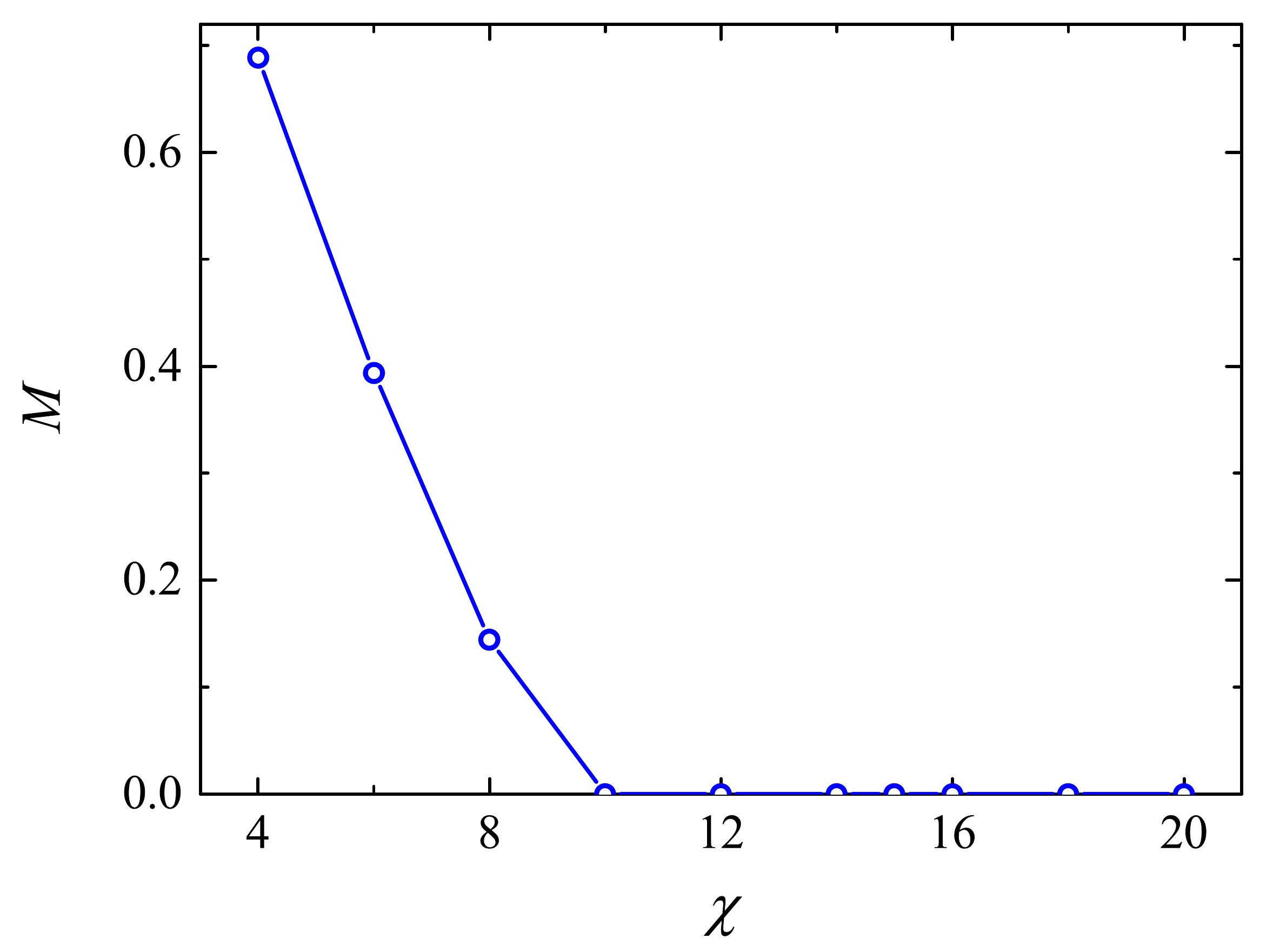}
\end{center}
  \caption{Magnetization $M$ of the spin-$2$ simplex solid state on a Kagome lattice obtained using the NTN method.
  }
\label{SSS}
\end{figure}

Below we use the infinite time-evolving block decimation (iTEBD)\cite{iTEBD1} method to demonstrate how to contract an RTN.
Similarly to the classical statistical model, $\langle \Psi | \Psi \rangle$ can be expressed as a trace over a product of the transfer matrix $T$, and the contraction of this RTN is equivalent to solving the dominant eigenvalue problem in the thermodynamic limit.
For the RTN defined by Eqs.~(\ref{eq:RTN1}) and (\ref{eq:RTN2}), $T$ itself is the product of four sub-transfer matrices, $T_1$, $T_2$, $T_3$, and $T_4$, whose graphical representations are shown in Fig.~\ref{ReducedTN}.
The iTEBD is used to solve the dominant eigen-problem of $T$ using an MPS,
\begin{equation}
|\Phi\rangle = \sum_{\{\sigma_i\}}\mathrm{Tr}(...W[\sigma_i]...)|\cdots\sigma_i\cdots\rangle , \label{eq:MPS1}
\end{equation}
where $W[\sigma_i]$ is a matrix of dimension $\chi$.
The trace sums over all the virtual bond states, and the summation runs over all configurations of $\{ \sigma_i \}$.
A graphical representation of this MPS is shown at the bottom of Fig.~\ref{ReducedTN}.
As the transfer matrix is translation invariant, we assume the local matrix $W$ to be site independent.
Initially, $W$ can take a random input.
It is then updated by successively and repeatedly multiplying the transfer matrices, $T_i$ ($i=1,\cdots , 4$), until it becomes converged.
After each multiplication, the bond dimension of $W$ becomes $\chi D^2$.
To maintain the iteration, we have to take a set of singular value decompositions to canonicalize the MPS~\cite{TEBD2} and then truncate the bond dimension of $W$ back to $\chi$.
The cost of performing singular value decomposition scales approximately as $O(\chi^3D^6)$ in computational time and $O(\chi^2D^4)$ in memory.
To contract the RTN accurately, $\chi$ is found to be at least of order $D^2$, as shown in Figs.~\ref{SSS} and \ref{Energy} as well as in Ref.~\onlinecite{Lubasch2014}.
Thus the computational time and memory costs scale at least as $O(D^{12})$ and $O(D^{8})$, respectively.
This high cost, as mentioned, has limited the largest accessible $D$ to around 13 with the currently available computer resources \cite{PESS2014}.

\section{The NTN method}
\label{sec:NTN}

Our solution to the problem encountered in contracting an RTN is to shift the top-layer by half a unit cell along the horizontal direction before compressing the double-layer tensor network into a single-layer one.
By doing this, we obtain a single-layer tensor network, i.e., an NTN, in which the local tensors in the top and bottom layers are intersected and connected by the physical bonds.
The major advantage of this method is that $D$ is not squared.
The bond dimension of the NTN, as illustrated in Fig.~\ref{fig:NTN}(a), is at most $D$ (the dimension of the physical bond $d$ is generally smaller than $D$), which is significantly smaller than the bond dimension of the RTN.
The price we have to pay is the doubling of the lattice size.
To contract the NTN using the TEBD (or iTEBD), we find that a relative larger $\chi$ is also needed in order to get a converged result in comparison with the RTN method.
However, as discussed below, the overall cost of this method is much lower than that of the RTN one.

In the NTN, a bond in the original top layer will intersect with one of the bonds in the original bottom layer [Fig.~\ref{fig:NTN}(a)].
If the iTEBD is directly applied to the NTN, each pair of these intersecting bonds will swap two local tensors in the MPS, which is difficult to handle.
To solve this problem, we introduce a swap gate $P$ for every pair of intersecting bonds, as illustrated in Fig.~\ref{fig:NTN}(b):
\begin{equation}
  P_{ijkl}=\delta_{ij}\delta_{kl} .
  \label{eq:P}
\end{equation}
The resulting NTN is shown in Fig.~\ref{fig:NTN}(c).
It is composed of $\{A, B, A^{\ast}, B^{\ast}, P\}$ tensors whose bond dimensions are equal to $D$ or $d$ instead of $D^2$.
The transfer matrix $T$ is now the product of six subtransfer matrices, $T_i$ ($i=1\cdots 6$),  two of which, i.e., $T_2$ and $T_5$ in Fig.~\ref{fig:NTN}(c), contain only the swap tensors $P$.

Similarly to the RTN, contraction of the NTN with the iTEBD is used to solve the dominant eigenproblem of the transfer matrix.
Corresponding to the nested structure of the NTN, the MPS is now defined by
\begin{equation}
  |\Phi\rangle = \sum_{\{\cdots\sigma_i\sigma_j\cdots\}}\mathrm{Tr} (\cdots W^a[\sigma_i]W^b[\sigma_j]\cdots) |\cdots\sigma_i\sigma_j\cdots\rangle    ,
  \label{eq:BMPS2}
\end{equation}
where $W^a[\sigma_i]$ and $W^b[\sigma_j]$ are two different local matrices of virtual bond dimension $\chi$.
To canonicalize the MPS, a set of singular value decompositions for matrices of dimension $\chi D$ should be done.
In this case, the leading computational time and memory costs scale as $O(\chi^3D^3)$ and $O(\chi^2D^2)$, respectively.
Again, to obtain a converged result, $\chi$ should be at least of order $D^2$, as mentioned.
Thus the costs scale approximately as $O(D^9)$ in computational time and $O(D^6)$ in memory.

\section{Results}\label{sec:result}

In Ref.~\onlinecite{PESS2014}, the PEPS was generalized to a PESS in which the many-body entanglement in a simplex, e.g., building blocks of a lattice, is emphasized. A triangle is the smallest simplex of the Kagome lattice.
If we take each triangle as a simplex, the corresponding PESS wave function is denoted the 3-PESS, where the prefix 3 represents the  number of sites at each simplex. Below we present the results obtained using the NTN method for the ground state of the Heisenberg model in the 3-PESS representation defined on a Kagome lattice \cite{PESS2014}.
The Kagome Heisenberg model is physically interesting because its ground state might be a quantum spin liquid \cite{SL1973, SL2010}.
This model has been extensively studied in recent years.
Some of the DMRG \cite{DMRGSL1, DMRGSL2, DMRGSL3}, coupled cluster expansion \cite{HOCC} and analytical Schwinger-boson mean-field calculation \cite{SBoson} suggest that its ground state is a gapped quantum spin liquid of $Z_2$ topology.
Other work, including analytical large-$N$ expansions \cite{LargeN}, variational Monte Carlo simulations \cite{VMCSL1, VMCSL2}, and, more recently, tensor renormalization group calculations \cite{LiaoPRL2017}, suggest that the ground state is a gapless spin liquid with U(1) symmetry and a Dirac spectrum of spinons.

Let us first consider two simple tensor-network states
whose exact or accurate results are known. One is the spin-2 simplex solid state~\cite{PESS2014, SSS} defined on a Kagome lattice.
This state can be represented as a $3$-PESS with a bond dimension of $D=3$ on the Kagome lattice, or, equivalently, a PEPS with the same bond dimension on the honeycomb lattice.
This state is the ground-state of the Hamiltonian defined by the sum of the $P_4$ projection operators introduced in Ref.~[\onlinecite{PESS2014}].
Both the energy $E$ and the magnetization $M$ of this state are exactly 0.
Using our method, we find that the ground-state energy is finite when $\chi \le 3$, but it becomes 0 within numerical error even when $\chi$ is as small as 4.
The magnetization, as shown in Fig.~\ref{SSS}, is also finite at small $\chi$, but it becomes 0 when $\chi$ is just above $D^2=9$.
It is clear that the energy converges more rapidly than the magnetization, although both are the expectation values of local physical quantities.

Another state used for testing is the resonating valence-bond state first introduced in Ref.~[\onlinecite{PESSRVB}] for the Kagome antiferromagnetic Heisenberg model.
This state can be represented as a 3-PESS of $D=3$ on the Kagome lattice as well.
Using our method, we find that the energy per site of this sate is about $-0.393124(1)$, in good agreement with the result obtained by the finite-size scaling~\cite{PESSRVB}, $-0.393123$.

\begin{figure}[t]
\begin{center}
\includegraphics[width=7.5cm]{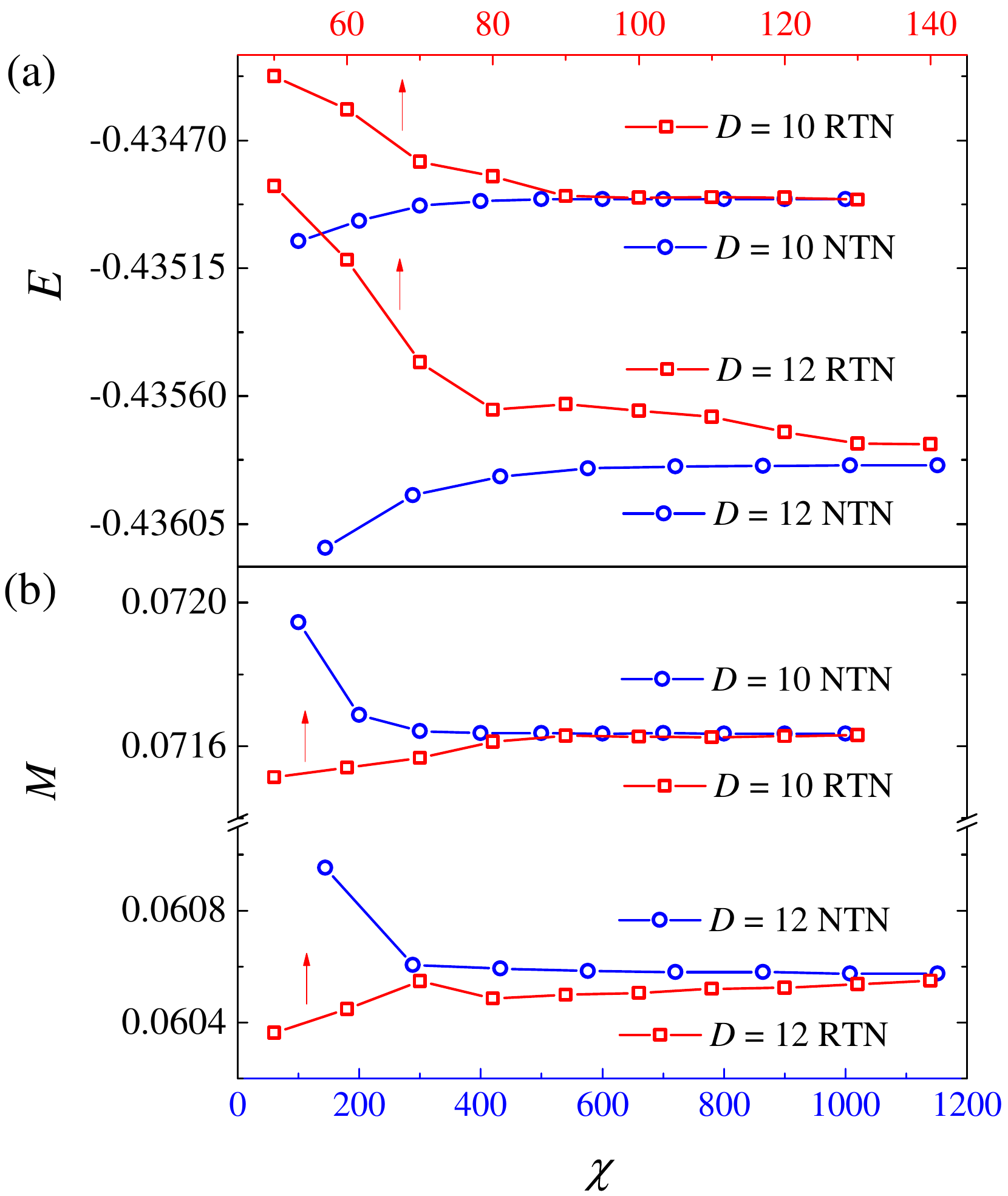}
\end{center}
  \caption{
  Bond dimension $\chi$ dependence of (a) the energy $E$ and (b) the magnetization $M$ obtained by the NTN (blue) and RTN (red) methods
  for the 3-PESS ground-state wave function of the $S=1/2$ Kagome Heisenberg model. The energy and magnetization can go either up or down with increasing $\chi$, since the expectation value calculations calculated with either the NTN or the RTN are not variational.
  }
\label{Energy}
\end{figure}

Now let us apply the method to study the physical properties of the ground state represented by the 3-PESS with a relatively large $D$ for the Kagome antiferromagnetic Heisenberg model.
This model is physically interesting because its ground state might be a quantum spin liquid \cite{SBoson, LargeN, DMRGSL2, KagomeSL4, DMRGSL3, LiaoPRL2017}.
The 3-PESS wave function is determined by the simple update method.

We first consider the case where $D$ is not too large, for example, $D=10$ or 12, so that both the NTN and the RTN can be reliably contracted by the iTEBD approximately.
We calculate both the energy and the magnetization for these wave functions. Figure \ref{Energy} compares the results obtained using the two contraction methods.
With an increase in the bond dimension $\chi$, the converged results obtained with these two methods agree with each other within numerical errors.
This shows that the value of $\chi$ that is needed to get a converged result is smaller for the RTN than for the NTN.
This is understandable because the entanglement between the two layers, i.e., $\langle\Psi|$ and $|\Psi\rangle$, is exactly considered in the RTN,
while in the NTN, this entanglement is just approximately treated because the physical bonds are truncated in the iTEBD iterations.
However, as revealed by the results for $D=12$, we can obtain even better converged results because much larger $\chi$ can be accessed by the iTEBD in the NTN calculation.

It should be pointed out that both the RTN and the NTN are approximate contraction methods. The variational principle is violated in the evaluation of the expectation values with either method, resulting from the truncation errors accumulated in the contraction of TNS by the TEBD or iTEBD method.
Thus the energy and magnetization obtained with this kind of method can go either up or down with increasing $\chi$.
Thus the only thing one should look for is the convergence, and a higher $\chi$ is the only figure of merit concerning the accuracy.

When $D$ becomes larger than 13, it is almost impossible to get a converged result by applying the iTEBD to the RTN due to the limitation of the values of $\chi$ that can be accessed. For example, for the $D = 20$ wave function, a $\chi$ as large as $D^2 = 400$ or more should be used to obtain converged results with the RTN method. In this case, the memory space needed to store the local tensors is formidably large, not to mention the computational time.

\begin{figure}[t]
\begin{center}
\includegraphics[width=7.5cm]{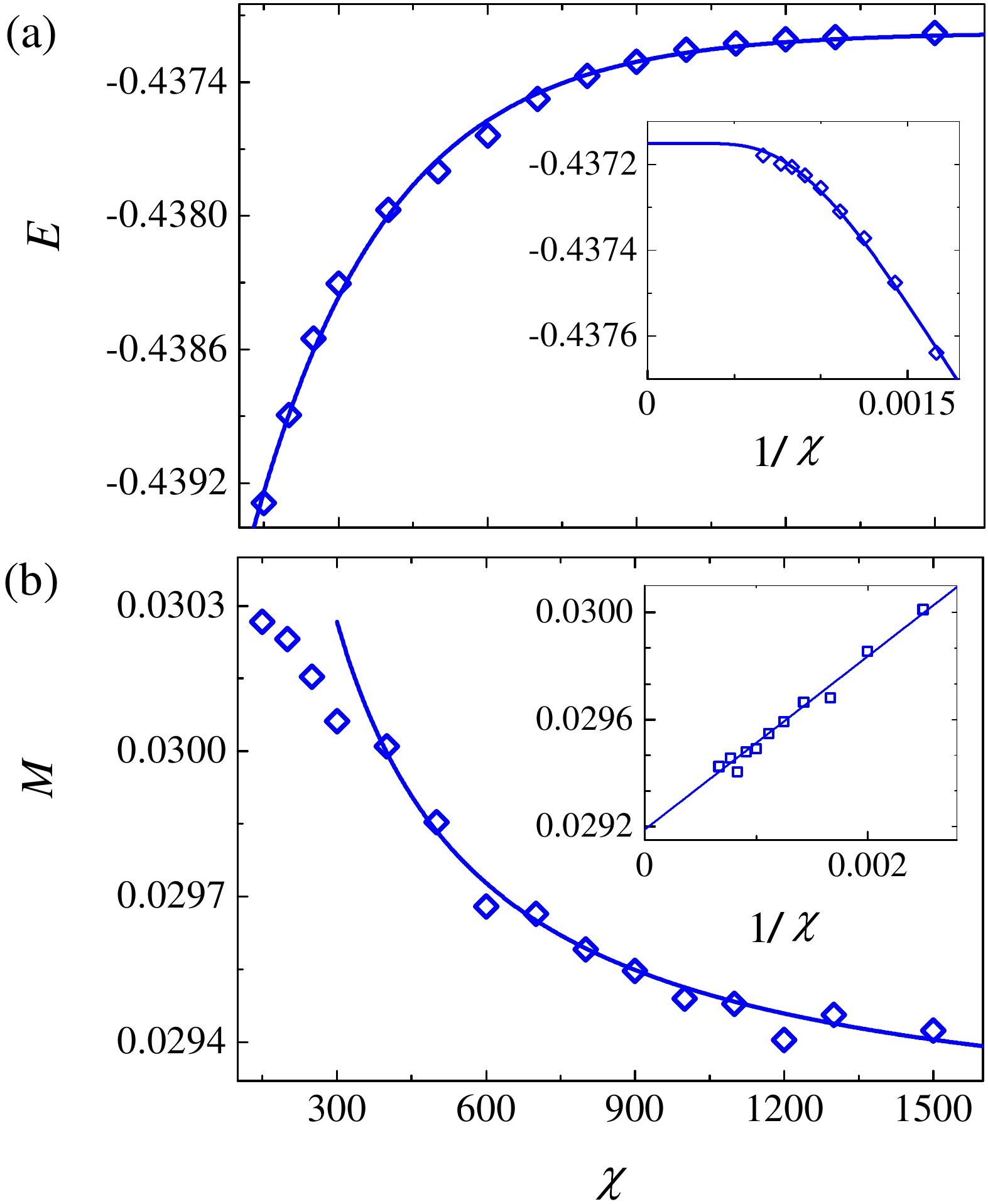}
\end{center}
  \caption{(a) Energy $E$ and (b) magnetization $M$ versus $\chi$ for the 3-PESS ground state of the Kagome antiferromagnetic Heisenberg model with $D=24$ obtained with the NTN method.  Insets: Exponential and linear fits to the small-$1/\chi$ data for $E$ and $M$, respectively.
  }
  \label{D24}
\end{figure}

Figure~\ref{D24} shows the $\chi$ dependence of the ground-state energy $E$ and the magnetization $M$ obtained using the NTN method from the 3-PESS wave function with $D = 24$ for the Kagome antiferromagetic Heisenberg model.
We find that the ground-state energy already becomes exponentially converged at $\chi \sim 1500$.
This means that we can get an accurate estimation of the ground-state energy by extrapolation.
The magnetization $M$ varies just algebraically in the same range of $\chi$, indicating that the results for $M$ are not fully converged even at $\chi = 1500$.
It shows that $M$ decreases with decreasing $1 / \chi$, hence we can get a lower bound of $M$ by extrapolating the results of $M$ with a polynomial function of $1/ \chi$.
The inset in Fig.~\ref{D24}(b) shows a linear fit of $M$ as a function of $1 /\chi$, from which we find the lower bound of $M$ to be about 0.0292 for the 3-PESS ground state of $D=24$.
The value of $M$ at $\chi = 1500$ is 0.0294, which can be taken as an upper bound of the magnetization for this state.
The true magnetization of this 3-PESS wave function is between these values.
The difference between these two values can be taken as the error for the magnetization, which is very small.

The accessibility of larger-$\chi$ calculations allows us to obtain not only a more accurate estimation of the ground-state energy, but also more useful information about low-lying excitations.
In a gapped system, the correlation length is finite, and the ground-state energy should converge exponentially with the bond dimension $D$ of the tensor-network state.
In a gapless system, however, the correlation length diverges and the ground-state energy should converge algebraically with $D$.
By calculating the 3-PESS wave function with $D$ up to 25, we recently found that the ground-state energy shows an algebraical dependence on $D$, suggesting that the ground state of this system is gapless~\cite{LiaoPRL2017}.

\section{Summary}\label{sec:summary}

We have proposed an efficient tensor-network contraction algorithm, based on a nested single-layer mapping of the double-layer tensor-network state shown in Fig.~\ref{fig:NTN}(c), to  evaluate the expectation values of tensor-network states.
It greatly reduces the computational cost in comparison with the conventional tensor-network contraction methods.
In particular, it reduces the computational time from $O(D^{12})$ to $O(D^9)$ and the memory cost from $O(D^8)$ to $O(D^6)$.
This method allows us to calculate accurately a TNS with $D$ up to 25.
By considering the symmetry of the Hamiltonian, one can block diagonalize the tensor-network states \cite{HHZhao2010, AOP2012, JWMeiSU2, SHJiangU1} to further reduce the computational cost, and extend the bond dimension to a larger value.

The method we propose works very generally.
It can be readily generalized to square or other two-dimensional lattice systems.
Furthermore, it can be combined with other tensor-network contraction algorithms, such as the corner transfer matrix \cite{CTM1} and transfer-matrix renormalization group \cite{TMRG1, TMRG2, TMRG3}, to calculate expectation values.
It can be also used in full update \cite{FU2008, FU2010} or variational \cite{PEPS2004} calculations of tensor-network states with larger $D$ values.

\section*{Acknowledgments}

We thank Bruce Normand for useful comments and suggestions and Hui-Hai Zhao and Yu-Zhi Liu for helpful discussions. This work was supported by the National Natural Science Foundation of China (Grant No.~11474331), and by the Ministry of Science and Technology of China (Grant No.~2016YFA0300503).

\bibliography{Ref}

\end{document}